\documentclass[twocolumn]{aastex62}

\usepackage{textcomp}
\usepackage{amsmath}

\submitjournal{ApJ}

\begin{document}

\title{Seeding the Formation of Mercurys:\\An Iron-sensitive Bouncing Barrier in Disk Magnetic Fields}

\correspondingauthor{Maximilian Kruss}
\email{maximilian.kruss@uni-due.de}

\author{Maximilian Kruss}
\affil{University of Duisburg-Essen, Faculty of Physics, Lotharstr. 1-21, 47057 Duisburg, Germany}

\author{Gerhard Wurm}
\affil{University of Duisburg-Essen, Faculty of Physics, Lotharstr. 1-21, 47057 Duisburg, Germany}

\begin{abstract}

The inner part of protoplanetary disks can be threaded by strong magnetic fields. In laboratory levitation experiments, we study how magnetic fields up to 7 mT influence the aggregation of dust by observing the self-consistent collisional evolution of particle ensembles. As dust samples we use mixtures of iron and quartz in different ratios. Without magnetic fields, particles in all samples grow into a bouncing barrier. These aggregates reversibly form larger clusters in the presence of magnetic fields. The size of these clusters depends on the strength of the magnetic field and the ratio between iron and quartz. The clustering increases the size of the largest entities by a factor of a few. If planetesimal formation is sensitive to the size of the largest aggregates, e.g. relying on streaming instabilities, then planetesimals will preferentially grow iron-rich in the inner region of protoplanetary disks. This might explain the iron gradient in the solar system and the formation of dense Mercury-like planets.

\end{abstract}

\keywords{protoplanetary disks --- planets and satellites: formation --- magnetic fields}

\section{Introduction}

Mercury is somewhat exceptional in the solar system as it has a particular interior structure with a rather large iron core \citep{Spohn2001, Hauck2013, Margot2018}. It is challenging to generate a Mercury-like planet in \textit{N}-body simulations from normal planetesimal distributions \citep{Lykawka2017}. A number of more selective ideas on its formation have been suggested. Current ideas on planetary evolution more generally include the possibility of evaporation of atmospheres \citep{Owen2017, Jin2018, Persson2018}. As an extreme case it has been suggested that evaporation might also include the mantle \citep{Cameron1985}. Removal of the mantle by a large impact event has also been proposed as an option \citep{Benz1988, Stewart2013, Asphaug2014}. Relying more on disk processing is photophoretic sorting at the edge of protoplanetary disks, which selectively removes silicates to the outer disk \citep{Wurm2013, Cuello2016}. Other mechanisms include inward drift of interplanetary dust particles (IDPs) from the outer solar system to change the local composition at Mercury's orbit \citep{Ebel2011} or magnetic erosion in the disk's magnetic field \citep{Hubbard2014}. The latter was motivation for this work, though we note that the mechanism studied here is quite different. With volatile elements measured to be rather high by the \textit{MESSENGER} mission not all scenarios might be viable \citep{Peplowski2011}.

The formation of an iron-rich planet might not be a singular event that happened only in the solar system. The extrasolar database now contains a number of planets that are rocky \citep{Marcy2014}. Among these are very dense ones as, e.g., noted by \citet{Rappaport2013} and \citet{Sinukoff2017}. \citet{Santerne2018} report the discovery of a Mercury-like exoplanet the size of the Earth. There are also multiple-planet systems by now harboring a denser planet \citep{Guenther2017}. 

In general, there is a trend in the solar system of decreasing iron content with radial distance to the Sun at least for the inner part. Meteorites, which sample asteroids, are depleted in iron compared to solar values \citep{Trieloff2006}. This includes oxidized iron, e.g. bound in silicates, but also metallic iron being present as pure iron-nickel grains within meteorites \citep{Trieloff2006}. The uncompressed densities of terrestrial planets also hint at a radial gradient with Mars being less dense than the other inner terrestrial planets, Mercury being the extreme \citep{Balogh2002}.

We will not discuss the pros and cons of the different formation mechanisms proposed for Mercury further but add the potential of a "new" one - magnetic aggregation. In principle, the idea of magnetic aggregation is not really new. The formation of large aggregates or nets of magnets has, e.g., been studied by \citet{Nuth1994}, \citet{Dominik2002}, and \citet{Nuebold2003} and is also a topic of current research in the more general sense of granular media, e.g., by \citet{Koegel2018}. In all of these cases, the particles are permanent magnets, though with given dipoles that might be oriented in random directions, leading to attraction or repulsion but preferring states of low energy, eventually. This is a different premise from our approach.

While all ferromagnetic minerals might have some residual magnetization, without an external magnetic field this is usually far from being saturated. Therefore, magnetization might strongly be enhanced in an external magnetic field. How aggregation is influenced by magnetic fields has been the subject of research in the fields of colloidal suspensions including, e.g., ferrofluids and magnetorheological fluids \citep{Lowen2008, deVicente2011}. Ferrofluids that consist of suspended magnetic nanoparticles tend to build chainlike structures due to magnetic dipole-dipole forces, whereas Brownian motion and interchain interaction may disperse such structures again. However, the stronger the magnetic field is, the stronger and more viscous the particle chains get \citep{Odenbach2002}. This is also the case when increasing the size of the suspended particles \citep{Gans2000}. \citet{Osipov1996} and \citet{Mendelev2004} derived theoretical distributions of the lengths of grown chains in dependence of the field strength. \citet{Zubarev2007} extended these considerations, taking into account non-identical constituents of the chains. If the applied magnetic field is not static but alternating, it can also be used to manipulate the size and shape of magnetic colloidal suspensions \citep{Snezhko2011,Snezhko2016}. The subject of this work is how this fits into the context of planet formation. How does magnetic aggregation help in forming planetesimals, especially Mercurys?

First, significant magnetic fields are needed. These exist in inner regions of protoplanetary disks but are hardly accessible for direct measurements. Future Atacama Large Millimeter/submillimeter Array (ALMA) observations could potentially provide more insight into the question of magnetic field orientations and strengths \citep{Bertrang2017}. Based on simulations, magnetic fields are estimated to be as large as 1\,mT at the inner disk edge, decreasing with radial distance \citep{Dudorov2014, Brauer2017}. \citet{Wardle2007} even gives a maximum estimation of several 10\,mT provided by equality of magnetic and thermal pressure in the midplane.
\citet{Donati2005} find 100\,mT at the inner edge of the FU Ori accretion disk.
 Therefore, the obvious idea is that aggregates with high metallic iron content might grow larger within these external magnetic fields. This is in analogy to aggregation of permanent magnets but with a preferred orientation and with attractive dipole forces being much stronger.

We see the importance of such aggregation as filling a current gap in planetesimal formation. This is set if aggregates have grown to (sub)millimeter size. At this stage, collisions are energetic enough to restructure very porous aggregates or aggregates of low fractal dimension. Frequent collisions therefore compact dust aggregates to a maximum at this point with filling factors of around 30\,\% \citep{Weidling2009, Teiser2011a}. After this compaction process, aggregates have lost their capability of further restructuring. Therefore, they have lost the option of dissipating collision energy in large quantities. Getting ever more elastic they only bounce off each other now \citep{Kelling2014, Kruss2016, Kruss2017}. \citet{Zsom2010} introduced this growth barrier as the bouncing barrier. Concentration by dust traps and streaming instabilities might eventually lead to a gravitational collapse of a cloudlet \citep{Youdin2005, Johansen2014, Simon2016}. Streaming instabilities need a minimum particle size though \citep{Bai2010, Drazkowska2014}. Therefore, if magnetic aggregation can shift the bouncing barrier, planetesimals might grow more easily. If growth is biased to iron-rich matter, inner disk planetesimals might be iron-rich and provide the building material for Mercury-like planets, eventually. In detail, being sensitive to magnetic fields, the radial decrease comes with a natural radial iron gradient in larger bodies.

The strongest form of magnetic aggregation will occur for ferromagnetic materials, metal iron being at the forefront, though some iron oxides also qualify. At the Curie temperature of about 1000\,K iron becomes paramagnetic. This limits magnetic aggregation to a region outside of the the 1000\,K line. This limit is also tempting if magnetic aggregation triggers growth in general, as recent work indicates a general dearth of terrestrial planets being hotter than 1000\,K \citep{Demirci2017}.

We leave these kinds of follow-up questions for future work. Here, we concentrate for the first time on the question of whether aggregation in a magnetic field changes the aggregation process significantly enough to have implications for the process of planet formation.

\section{Experiments}

\begin{figure}[h]
	\includegraphics[width=\columnwidth]{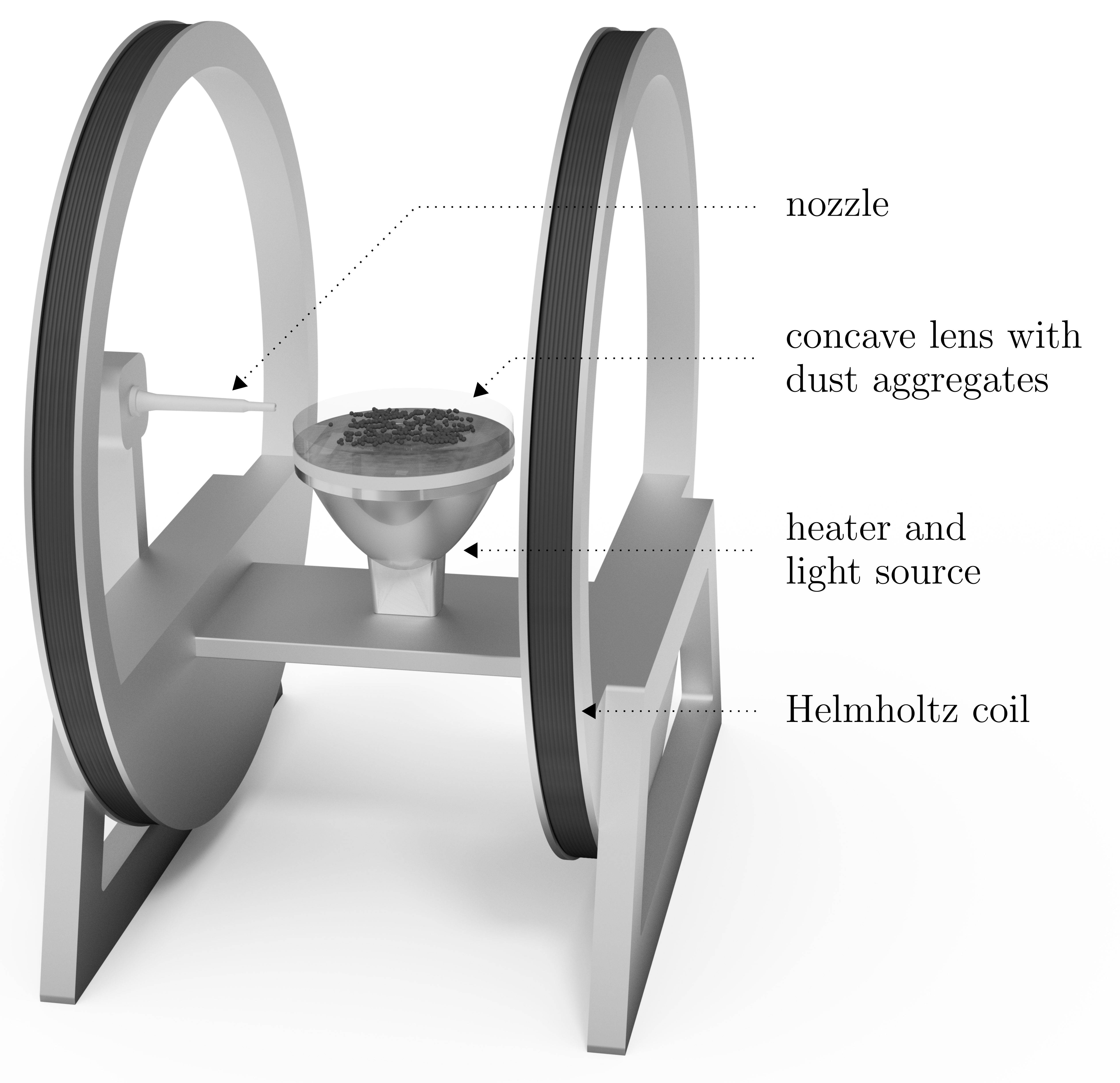}
	\caption{\label{fig.setup}Schematic of the experiment.}
\end{figure}

In recent years, the bouncing barrier has been studied in an experiment where aggregates are levitated by thermal creep on a hot surface \citep{Jankowski2012, Kelling2014, Kruss2016, Kruss2017, Demirci2017}. We use this setup here as the primary setup and add a pair of Helmholtz coils to provide a homogeneous magnetic field. A schematic of the experiment can be seen in Figure~\ref{fig.setup}.

Aggregates of dust are placed onto a heater and levitate at low ambient pressure. The aggregates then slowly collide with each other. For details of the levitation concept, we refer to \citet{Kelling2009} and \citet{Kelling2014}. However, for readability of this paper, we would like to summarize the major features as follows. The dust aggregates are heated from below by a halogen bulb to a temperature of about 800\,K while they are free to cool by thermal radiation on their top. This results in a temperature gradient over the aggregate. If the experiment is placed in a vacuum chamber with low ambient pressure, gas flows from above the aggregates through the pores of the aggregates toward their bottom. This is called thermal creep and requires the mean free path of the gas molecules to be comparable to the pore size \citep{Koester2017, Schywek2017, Steinpilz2017}. Therefore, if the pressure is reduced to around 10\,mbar, thermal creep leads to an overpressure below the aggregates that then lift off and hover over the surface. They randomly move around driven by some asymmetries in the thermal creep flow or excited by the occasional flow of gas through an air nozzle. Furthermore, the levitation platform, which consists of a glass lens, has a concave shape keeping the aggregates in the central part. This way collisions between the aggregates occur at typical collision speeds of mm\,s$^{-1}$ to cm\,s$^{-1}$. The motion and evolution are observed by a camera from above at a frame rate of 100\,fps. Since the halogen bulb also acts as a light source, brightfield observations proved to provide the best contrast. In the images, the aggregates appear as shadows.

Helmholtz coils are added to the setup generating a homogeneous field in the central part where the aggregates levitate. Thus, the field only magnetizes the grains but does not induce significant motion of an individual dipole. We measured and calibrated the field in the center for a given current by using a Hall effect sensor. The maximum field that can be achieved with the current setup is 7\,mT.

\subsection{Dust samples}

\begin{figure}[h]
	\includegraphics[width=\columnwidth]{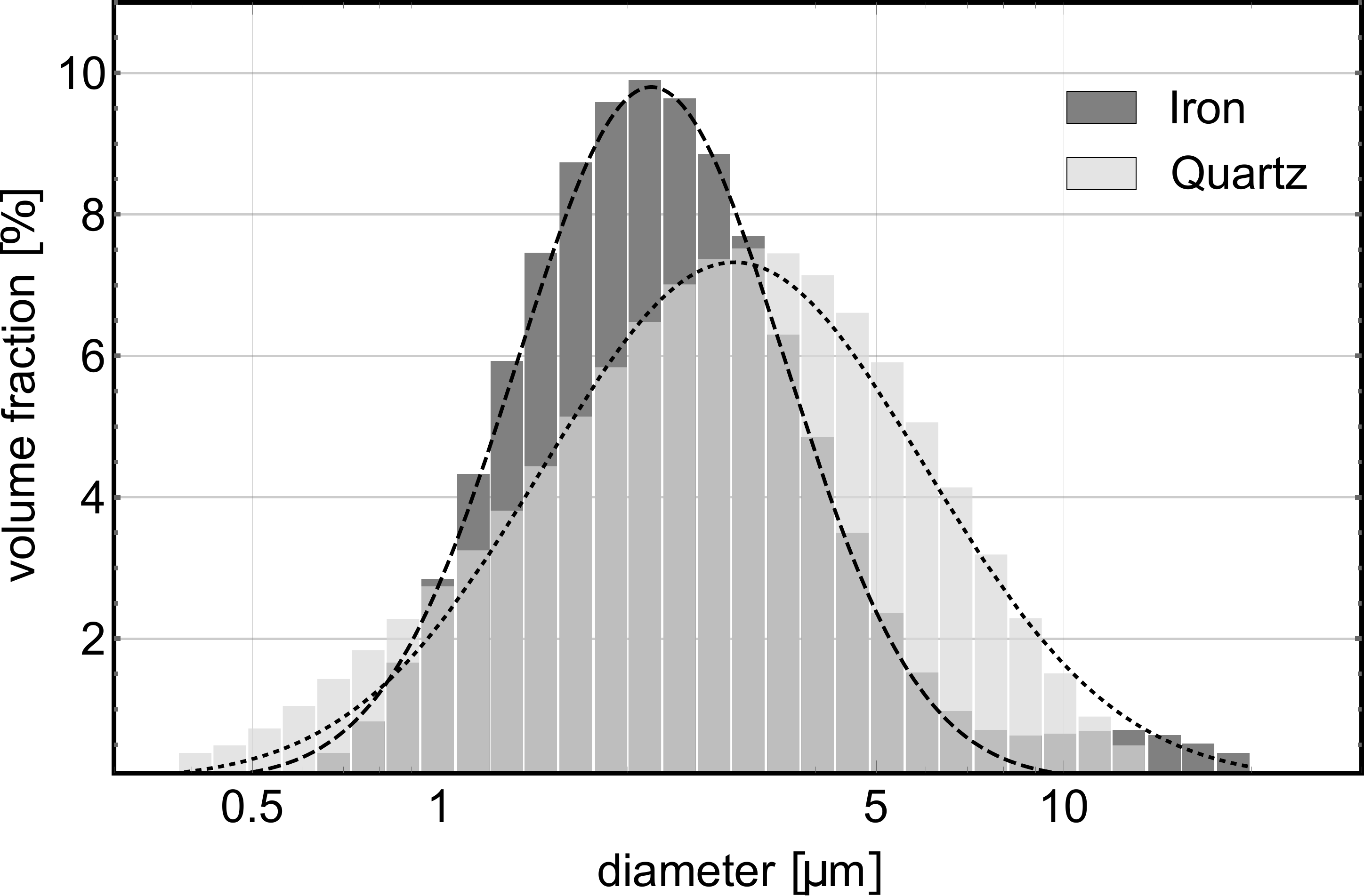}
	\caption{\label{fig.size}Size distribution of grains composing the aggregates. The data were measured by a commercial device based on light scattering (Malvern Mastersizer 3000). Log-normal distributions are included that peak at 2.2\,\textmu m and 3.0\,\textmu m, respectively.}
\end{figure}

\begin{figure}[h]
	\includegraphics[width=\columnwidth]{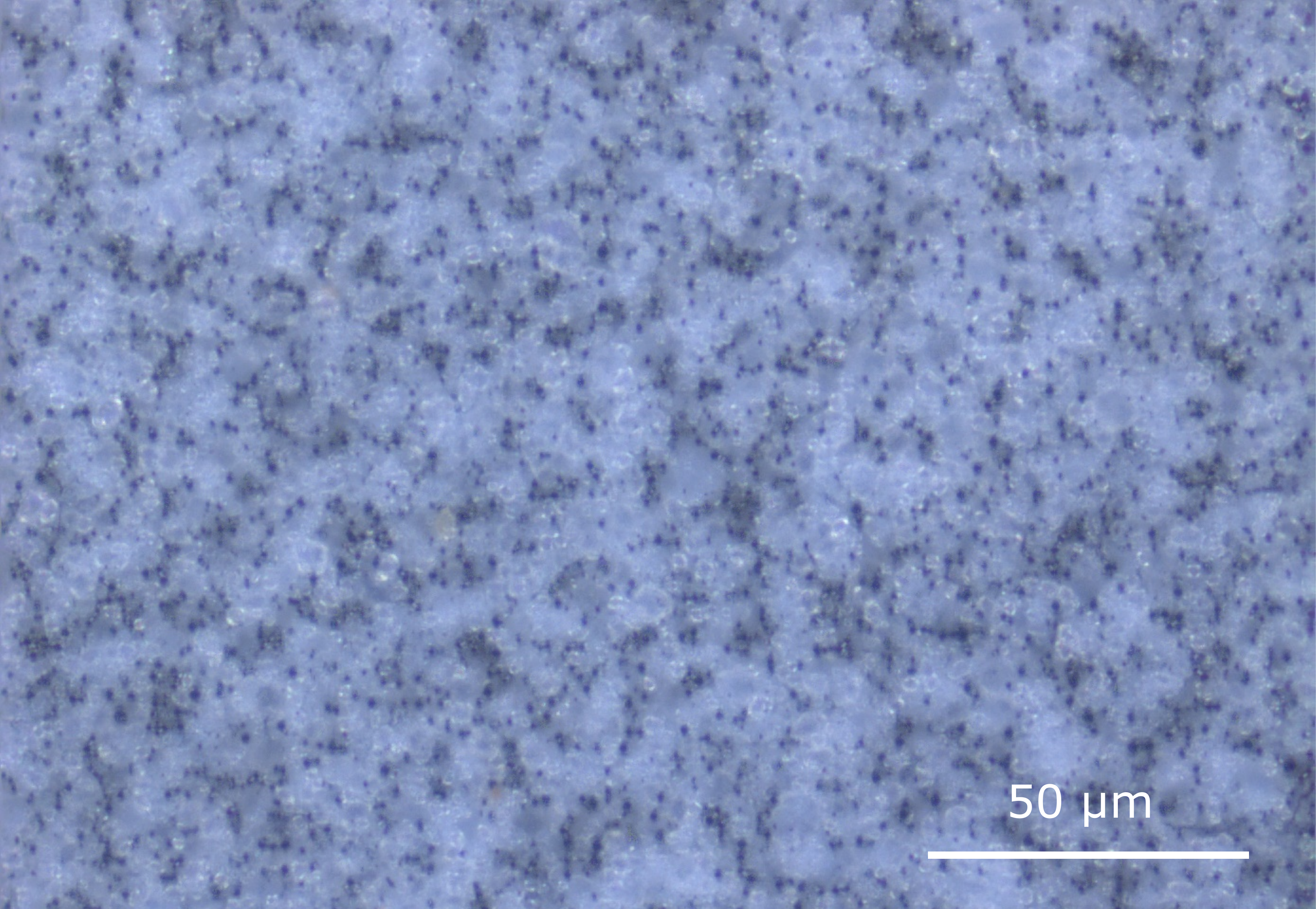}
	\caption{\label{fig.microscope}Microscope image of iron-rich aggregates premixed with quartz in a 1\,:\,1 mass ratio.}
\end{figure}

\begin{figure}[h]
	\includegraphics[width=\columnwidth]{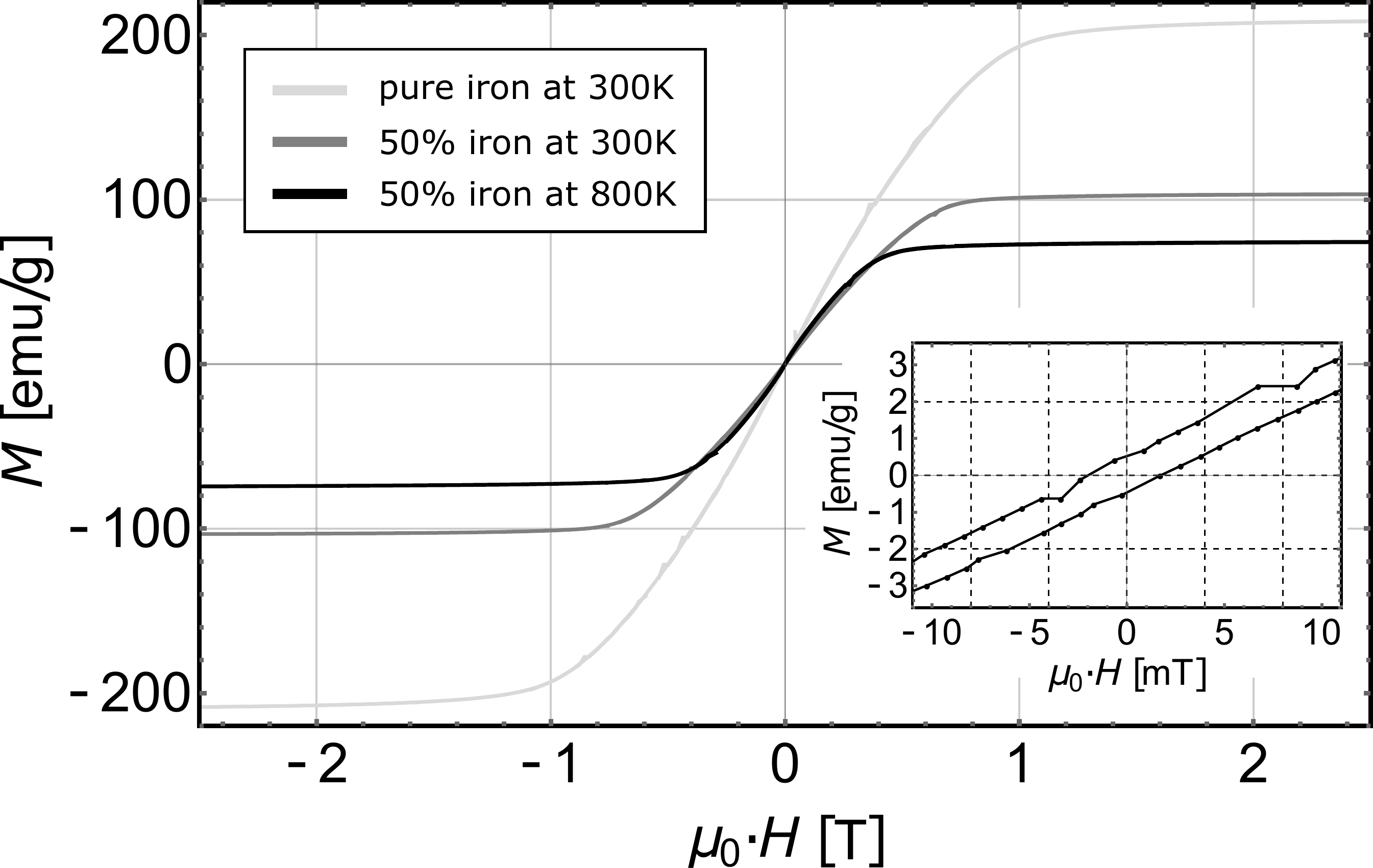}
	\caption{\label{fig.magnetization}Magnetization curves measured by a vibrating sample magnetometer for pure iron dust and premixed dust. The latter was also characterized at 800\,K, which is the temperature reached in the experiments. The x-axis shows the externally applied magnetic induction.}
\end{figure}

Basic dust samples used for the experiments were iron and quartz with average grain sizes of around 2--3\,\textmu m as can be seen in Figure~\ref{fig.size}. Pure iron has a high density and a high thermal conductivity that do not allow levitation of pure iron dust samples. We therefore premixed iron and silicate in different mass ratios to enable the particles to move freely. Out of this sample the starting aggregates were placed onto the heater by sieving them through a 180\,\textmu m mesh. 
An example of a microscope image of these mixed iron-rich aggregates is shown in Figure~\ref{fig.microscope}. The iron grains are evenly distributed within the sample.

We consider these mixed but iron-rich aggregates as stable as they do not fragment within the limits of our spatial resolution. Earlier work by \citet{Kruss2016} showed that a small amount of mass transfer at contact during a collision is possible. However, over the timescales of our experiments we consider this to be insignificant.

When investigating magnetic aggregation, it is essential to know the magnetic properties of the material. Figure~\ref{fig.magnetization} shows the magnetization curves of the used samples in an external magnetic field. Each measurement was performed with around 10\,mg of the respective sample in powder form. The powder was filled into a cylindrical sample container with a diameter of 3\,mm and a height of less than 1\,mm. The saturation magnetization of the premixed dust with an iron to quartz mass ratio of 1:1 is around 104\,emu\,g$^{-1}$. This is in agreement with the saturation magnetization of pure iron dust, which is 211\,emu\,g$^{-1}$. When heating the sample to 800\,K, as was done in the experiments, the magnetic properties change, e.g., the saturation magnetization is decreased. In the range of magnetic fields that are relevant for this study, the magnetization is far from being saturated, which can be seen in the inset of Figure~\ref{fig.magnetization}. At a magnetic field of 7\,mT the magnetization of premixed dust at 800\,K reaches 2.5\,emu\,g$^{-1}$ which is around five times the value of the remanence.

\section{Results}

\begin{figure}[h]
	\includegraphics[width=\columnwidth]{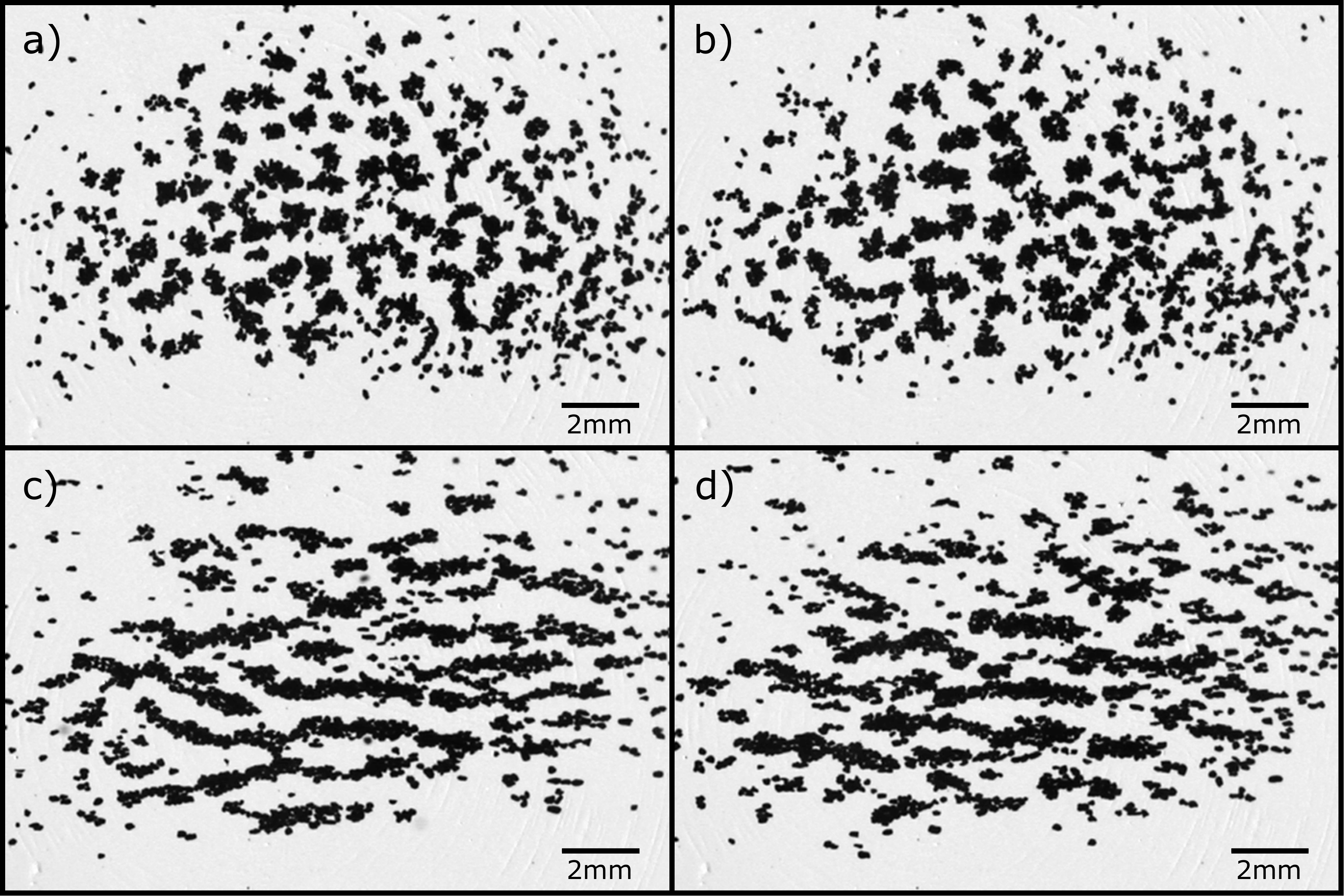}
	\caption{\label{fig.bilder}Aggregates formed without a magnetic field at $t_\text{off}$ (a) and $t_\text{off}\,+\,0.1$\,s (b) compared to aggregates formed with a magnetic field of 7\,mT applied at $t_\text{on}$ (c) and $t_\text{on}\,+\,0.1$\,s (d). The last two images were recorded 3\,s after the application of the field.}
\end{figure}

\begin{figure}[h]
	\includegraphics[width=\columnwidth]{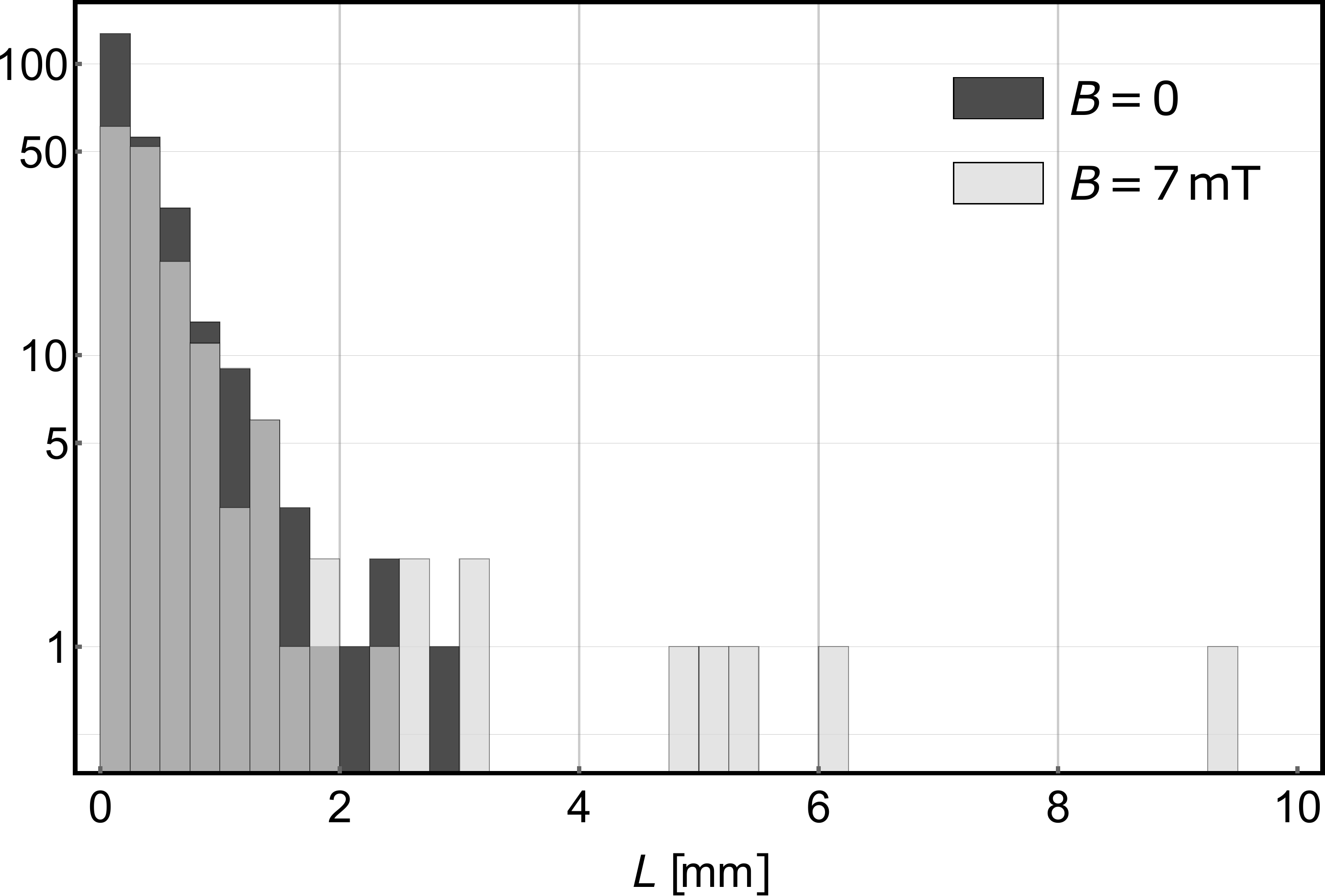}
	\caption{\label{fig.histogram}Histogram of the chain lengths of grown aggregates in the direction of the magnetic field. The distributions correspond to the ensembles shown in Figure~\ref{fig.bilder}~(a,c).}
\end{figure}

\begin{figure}[h]
	\includegraphics[width=\columnwidth]{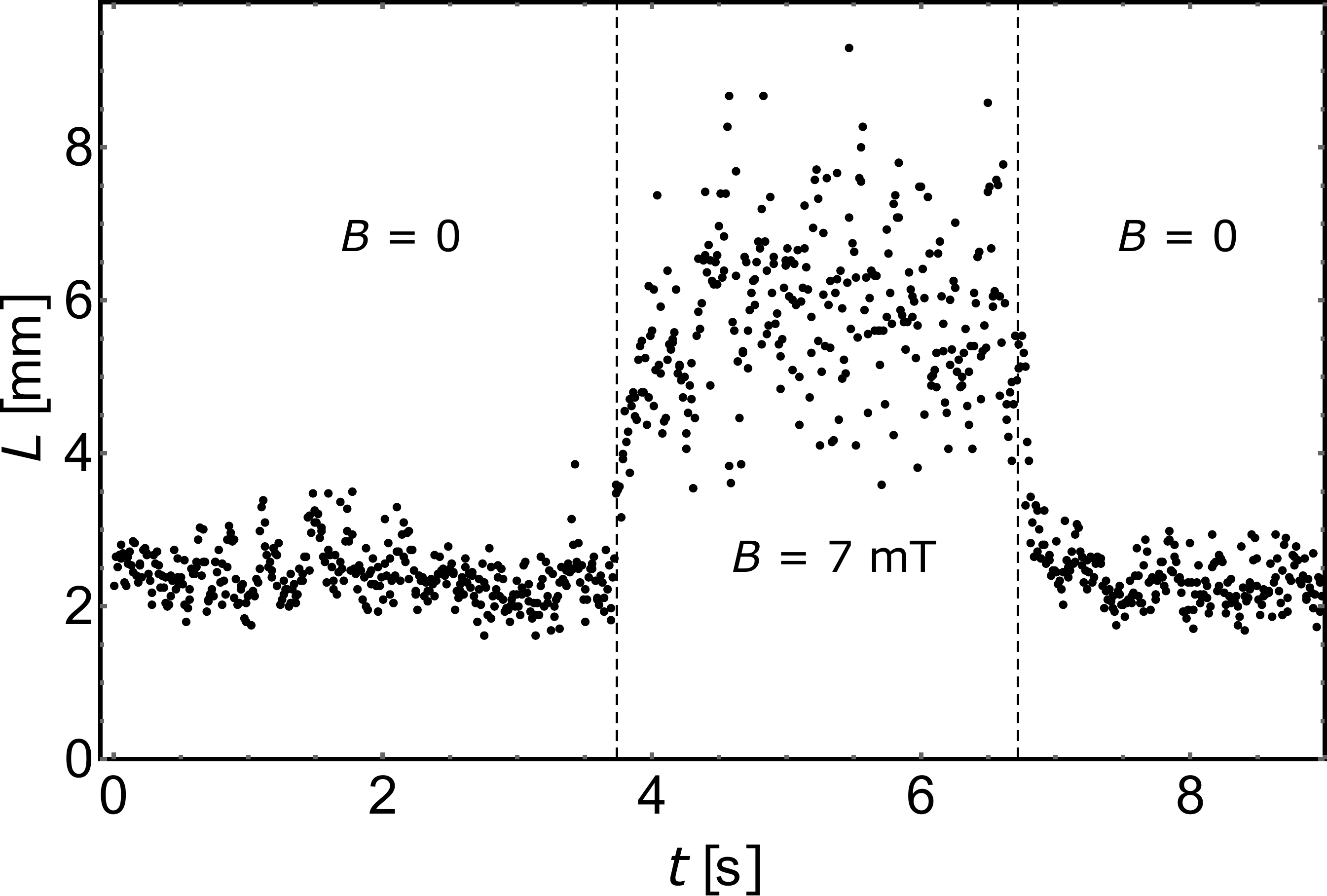}
	\caption{\label{fig.chainsize}Temporal evolution of the average chain length $L$ with the magnetic field turned on and off again.}
\end{figure}

Once the particles start to levitate, they are free to interact and stick to each other. Aggregates grow until they reach the bouncing barrier as was observed in former works. Figures~\ref{fig.bilder}(a),~(b) show those aggregates at the bouncing barrier without the external magnetic field. As soon as the magnetic field is switched on, the aggregates form particle chains in contrast to the results of experiments with permanent magnets mentioned above. If the magnetic field is turned off again, a significant remanent magnetization is left in the dust (see Figure~\ref{fig.magnetization}). However, the remanent magnetic dipoles are not aligned by an external field anymore but are orientated randomly. Therefore, the magnetic forces between the iron particles cannot compensate the forces acting during a collision. The chains disperse again readily, and aggregates return to bouncing at smaller size. This shows that the size of the aggregates or clusters of aggregates at the bouncing barrier strongly depends on the magnetic field. It should be noted that the ensembles of aggregates are highly dynamic. Due to the gas flows mentioned above, aggregates may change their size and shape within much less than a second and no steady state is reached, which is illustrated in Figures~\ref{fig.bilder}(b),~(d).

We determined the chain length of connected particles to measure the influence of the magnetic field as it is also a commonly used quantity in the related systems of ferrofluids and magnetorheological fluids \citep{Gans2000,Mendelev2004}. The images were binarized using Otsu's algorithm \citep{Otsu1979} so that the lengths of all aggregate chains in the direction of the magnetic field could be determined automatically. A histogram of the chain lengths is shown in Figure~\ref{fig.histogram}. Since the size of the experimental platform is limiting the amount of the used dust, most of the aggregates, especially those in the outer parts of the ensembles, cannot be considered as fully grown. In order to exclude these smaller particles, the following analysis relies on the average length of the five largest chains in every single frame. This reduction seems reasonable looking at typical distributions as in Figure~\ref{fig.histogram}. An exemplary evolution of the chain length can be seen in Figure~\ref{fig.chainsize}. The largest chains formed in the magnetic field are larger than those without magnetic field by a factor of up to 3. However, this is not a strict upper limit. The limited amount of the used dust also puts constraints on the maximum chain length at the highest applied fields.
 
\begin{figure}[h]
	\includegraphics[width=\columnwidth]{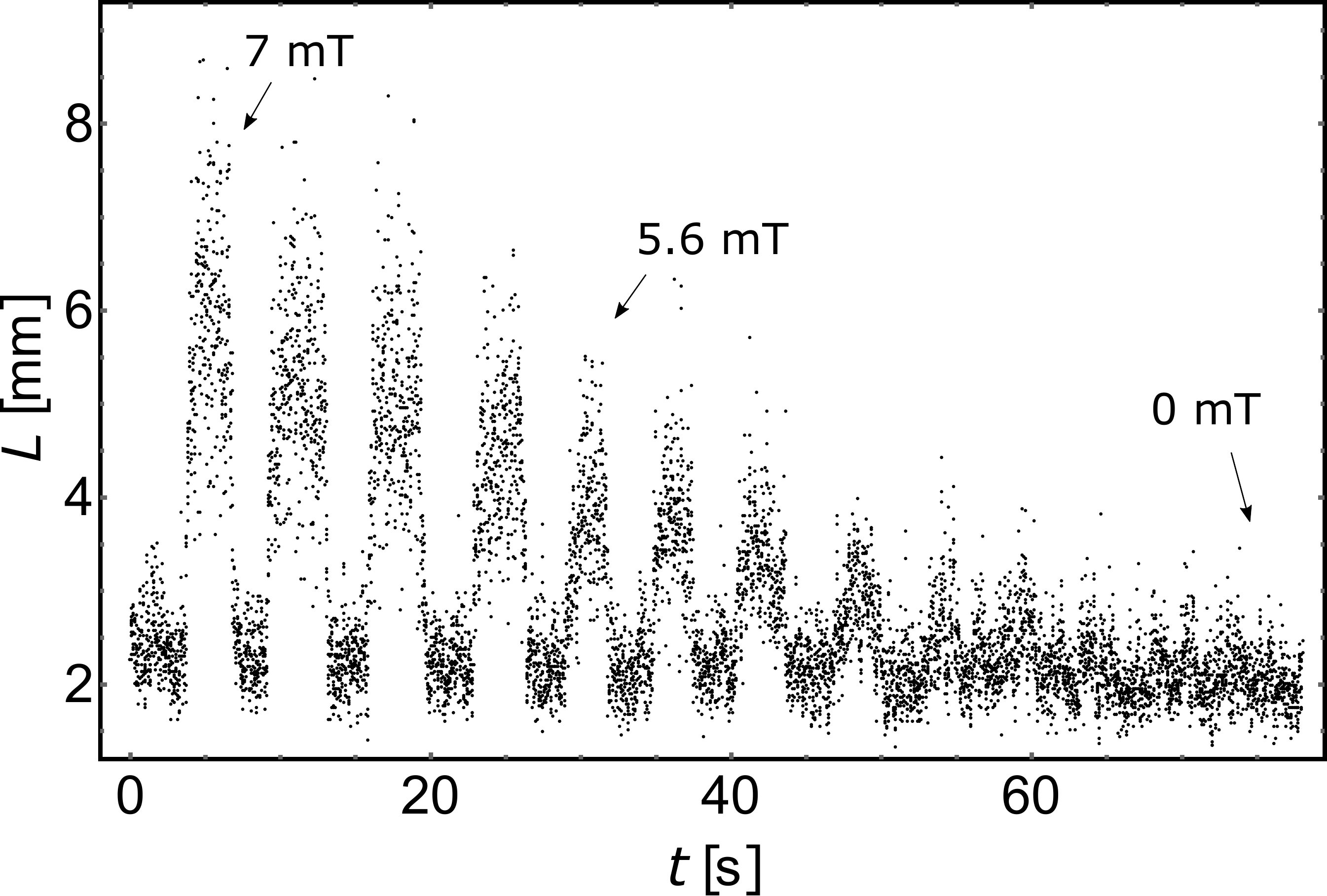}
	\caption{\label{fig.verlauf}Chain length $L$ changing with magnetic field being alternately turned on and off. The magnetic field is decreased stepwise from 7\,mT to 0.}
\end{figure}

\begin{figure}[h]
	\includegraphics[width=\columnwidth]{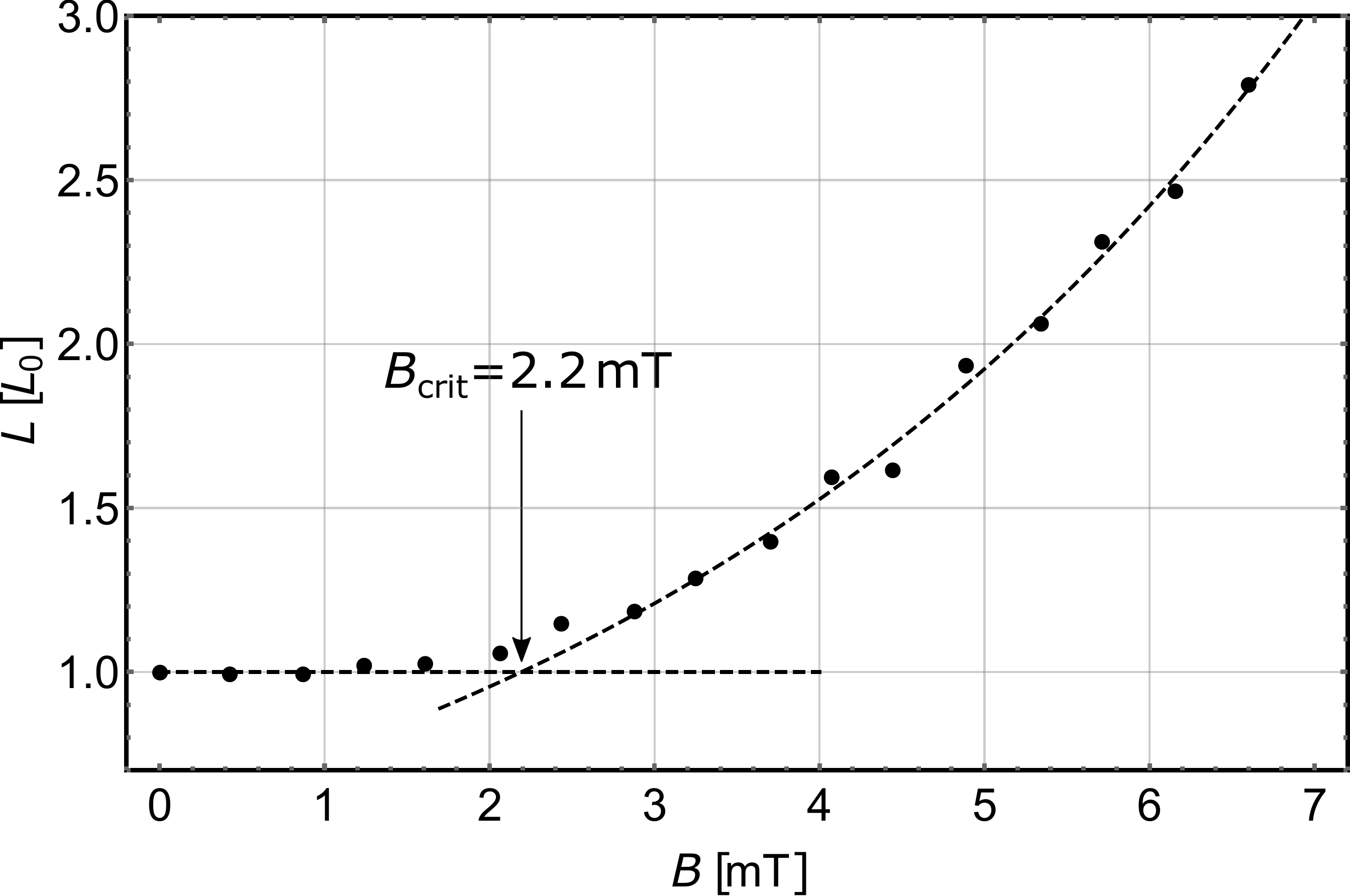}
	\caption{\label{fig.bcrit}Length \textit{L} of field-grown chains in units of no-field-grown chains $L_0$ over applied magnetic field. The data points are mean values of the chain sizes during the respective intervals.}
\end{figure}

To further quantify the effect of increasing chain length, we carried out measurements at different magnetic field strengths. Figure~\ref{fig.verlauf} shows the evolution of the chain size grown with a magnetic field of decreasing strength being turned on and off alternately. From this data we deduced the dependence of the mean chain length on the magnetic field, which is depicted in Figure~\ref{fig.bcrit}.

The dashed lines indicate two different regimes. At low magnetic fields the chain length remained unchanged, while at higher fields we used an exponential fit to model the increase in length. We note here that no physical model is underlying the exponential fit. However, it seems useful as a quantitative estimate for the influence of the magnetic field, especially for comparing different measurements. We define the intersection of the fits as the critical field $\textit{B}_\text{crit}$. This marks the strength of the magnetic field where chains start to form and where magnetic attraction is stronger than collisional repulsion and all disturbing effects. In the case of an iron to silicate ratio of 1\,:\,1, the bouncing barrier should be influenced by magnetic fields larger than 2.2\,mT.

\begin{figure}[h]
	\includegraphics[width=\columnwidth]{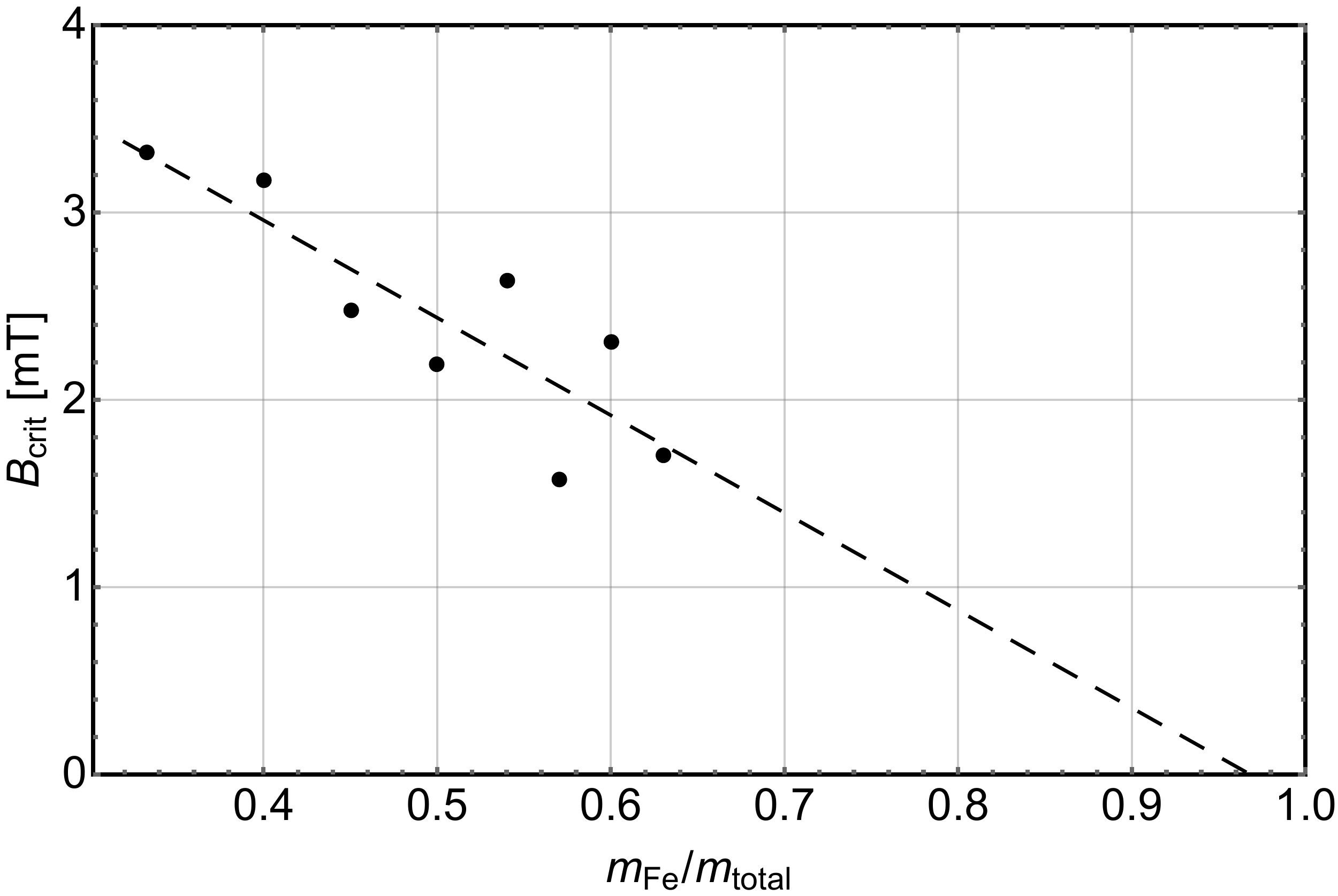}
	\caption{\label{fig.ironfraction}Critical magnetic field where chains start to grow in dependence of the iron mass fraction of the used aggregates. A dashed line is included to extrapolate the drop of $\textit{B}_\text{crit}$ toward higher iron content.}
\end{figure}

To see the effect of more iron-rich material, we systematically varied the iron content of the premixed dust aggregates. The result is shown in Figure~\ref{fig.ironfraction}. The minimum iron mass fraction leading to chain formation was 0.33. Above a value of 0.63 the iron content was too high to allow levitation of the aggregates. A dashed line is included to guide the eye toward higher iron content and to give an estimate of how pure iron particles would be influenced. Given a decrease of $\textit{B}_\text{crit}$ as extrapolated in Figure~\ref{fig.ironfraction}, pure iron aggregates would start forming chains -- under the experimental conditions -- well below 1\,mT.

\section{Caveats and Discussion}

The collisions of the aggregates are not perfect as there are still residual forces acting due to the levitation mechanism, and the collisions are essentially only 2d (see \citet{Kelling2014}).

We started with artificially, preformed aggregates of 180\,\textmu m composed of quartz and iron that interacted with each other. Our work does not explain how preferred iron aggregates are formed in the first place. It is likely that such a bias occurs early, but this requires research with mixed particle clouds that is not easily achieved. 

It is subject to debate how much metallic iron or ferromagnetic minerals are really available. We only detail here how a possible existence of pure iron grains would influence the size of aggregates.

The aggregates grow to larger aggregates by hit-and-stick collisions until reaching the bouncing barrier. \citet{Kruss2017} and \citet{Demirci2017} observed a similar evolution toward the bouncing barrier for quartz and basaltic samples.
With a magnetic field applied, we clearly observe effects of alignment of iron grains on aggregation. We do not fragment on an individual dust grain level here but on an aggregate level at the bouncing barrier. We consider this to be close to the situation of a protoplanetary disk for initial grain growth, where the bouncing barrier is a first limit, and fragmenting collisions in the sense of disrupting a compact aggregate are not possible yet as, e.g., being the core of magnetic erosion suggested by \citet{Hubbard2014}. This might be a viable selective process but only acting later. 

Magnetic fields in protoplanetary disks are predicted to be up to the mT level \citep{Dudorov2014, Brauer2017}. Based on the extrapolation presented above, this is well within the region where a bias in growth between iron or ferromagnetic minerals and silicates would occur. For different collisional settings, e.g. collision velocities, 3d collisions or varying primary grain sizes, the minimum magnetic field needed will be different. As we did not study any variations yet or try to optimize any parameter to promote magnetic aggregation but already find it to be of importance, it is likely that even lower magnetic fields might already have a significant impact for other settings. This is subject to future research.
In any case, the bouncing barrier can be shifted toward larger aggregates for iron-rich materials.

For further evolution of the aggregates, the Stokes number, which is the ratio of the gas-grain coupling time over the orbital time, is important. The coupling time depends on the mass to surface ratio. For compact aggregates this is linear in size, but it is different for elongated chainlike aggregates depending on their orientation relative to the particle-gas motion. If an aggregate moves along the chain direction, the mass to surface ratio increases much more strongly compared to the case where the chains would drift perpendicular to their axes. We currently consider this a minor detail at this stage, but it should be kept in mind that the relative directions or orientation of gas flow, particle motion, and magnetic field might influence the aggregation and later phases.

\section{Conclusion}

The bouncing barrier is an important evolutionary step in planet formation that prevents dust aggregates from growing larger than millimeter size. In the presence of a magnetic field, however, iron-rich aggregates might grow somewhat larger due to magnetic dipole-dipole interaction. Silicate aggregates are essentially excluded from this evolution step. Our experiments show that field strengths below 1\,mT could be sufficient to boost growth of chainlike clusters if they are mostly composed of iron. The exact evolution strongly depends on the iron fraction of the aggregates and the magnetic field applied. However, as our experimental conditions are in a range expected in a protoplanetary disk, the effects might be significant for planet formation. Biased magnetic aggregation of iron-rich material might allow these aggregates to grow into the range accessible to streaming instabilities.
Ultimately, magnetic aggregation might then provide the seeding for the formation of Mercury-like planets.

\section*{acknowledgements}
This project is supported by the DFG under grant WU~321/14-1.
We are grateful to J. Landers, B. Eggert, and H. Wende for providing the magnetization measurements and acknowledge support by DFG grants  WU~321/18-1 and WE~2623/19-1.
We also appreciate the very helpful reviews of the two referees.

\bibliography{bib}

\end{document}